# Spin reconstruction in quantum wires subject to a perpendicular magnetic field

Gilad Barak[1], Loren N. Pfeiffer[2], Ken W. West[2], Bertrand I. Halperin[1], Amir Yacoby[1]

[1]*Department of Physics, Harvard University, Cambridge MA, 02138, USA.*
[2]*Department of Electrical Engineering, Princeton University, Princeton, NJ, USA*

We study the effects of a perpendicular magnetic field on the spin and charge distribution across a quantum wire. Using momentum resolved tunneling between two parallel wires we measure the dispersion relation for different perpendicular magnetic fields. We find that as the magnetic field increases, a strip of spin polarized electrons separates in the cross section of the wire. We provide a quantitative description of the emerging structure within a Hartree-Fock approximation, showing that it results from exchange interaction between electrons in the wire. We discuss the relation of these results to the structure of Quantum Hall edge states.

Many particle systems often arrange themselves in complex ways in order to reduce the energy cost associated with inter particle interactions. A well known example is the quantum Hall effect that arises when a two dimensional electron gas (2DEG) is subject to a strong magnetic field which effectively quenches the kinetic energy of the particles, giving rise to correlated incompressible phases in the bulk [1]. At sample boundaries, the competition between the confining potential and Coulombic interactions leads to nontrivial spin and charge textures [2-4]. However, to date only the simplest form of edge reconstruction associated with a slow decrease in particle density near the edge [2] has been experimentally observed [5]. While this form of charge redistribution can take place on a relatively long spatial dimension, several hundreds of nm in width, other possible reconstructions are predicted to extend over only a few magnetic lengths ($l_B = \sqrt{h/eB}$), making them very difficult to observe. Here, $h$ is the Planck constant, $e$ the electron charge and $B$ the applied magnetic field.

Let us consider as a concrete example the depletion of a Landau level near the edge of a 2DEG, when both spin orientations of this level are populated in the bulk. The charge structure near the edge is determined by a competition between the confining potential, promoting a sharp drop of the charge density, and the Hartree (direct) interaction which favors expanding the distribution towards the system edge. Dempsey *et al.* have shown that this competition gives rise to a 2$^{nd}$ order phase transition, such that for steep confinement the charge occupation near the edge is expected to drop discontinuously, whereas for smoother confinement the occupation drop will be gradual [3]. Exchange interaction plays an important role in the later case, where it supports a step-like decrease in the occupation from the fully occupied bulk, to a half-occupied strip in which only one spin type is populated instead of a continuous decrease in the charge density. The physics responsible for such a reconstruction is based on the fact that adjacent electrons of equal spin gain from exchange interaction, promoting spin polarization and such effects as ferromagnetism and Hund's rule of maximum multiplicity. While this predisposition is usually contrasted by the added kinetic energy cost involved in polarizing the electron distribution, for systems under the QHE state, when the kinetic energy is quenched, it favors maximal local occupation of a Landau level. For example, this mechanism promotes the formation of stripe and bubble textures, in which domains of two integer filling factors are created instead of a homogeneous non-integer filling factor [6].

The Dempsey reconstruction [3], therefore, predicts the creation of a spin polarized strip of typical width $l_B$ at the edge, through a $2^{nd}$ order phase transition. Correspondingly, the dispersions of the two spin edge excitations become distinct near the Fermi energy. This prediction has been extended to the case of electrons confined to a finite width quantum wire subject to a perpendicular magnetic field [7-9]. In this work we provide the first direct measurement of the Dempsey reconstruction in a one dimensional GaAs quantum wire.

We use momentum resolved tunneling spectroscopy between two parallel quantum wires, to study the dispersion of electrons confined to one dimension under a perpendicular magnetic field. The wires are prepared by the cleaved edge overgrowth method [10]. Two parallel quantum wells (QW) are grown in a GaAs/AlGaAs heterostructure. The upper well is 20-25nm wide, the lower is 30nm wide and they are separated by a 6nm, 300meV high barrier. A modulation doping sequence is used, which renders the upper QW occupied by a two-dimensional electron gas (2DEG), while the lower well is unpopulated. The sample is cleaved inside the MBE chamber perpendicular to the initial growth direction, and another modulation doping sequence is grown on the cleaved edge, creating a triangular potential well along the edge (see Fig. 1a). This additional confinement binds electrons to the edge and leads to the creation of quasi-1D modes with a long backscattering mean free path of tens of μm. The upper system (US) is thus comprised of the 'upper quantum wire' (UW) having a few quasi-1D modes as well as a populated 2DEG. In the lower well, electrons populate 1D states creating a 'lower quantum wire' (LW). One metal gate is used to electrostatically deplete electrons from both the *US* and the *LW* beneath it, and a second gate depletes only the *US* while leaving the *LW* continuous, thus defining two junctions (Fig. 1b). The short 'source' junction bordered by the two gates can have a length of 6-40μm. The long 'drain' junction bordered by a gate and the sample edge is of millimeter length. Contacts are made to the *2DEG* in the source and drain regions, which is coupled to the *UW* effectively. In order for current to pass between the leads, electrons must tunnel from the *UW* to the *LW* at the source junction, followed by a second tunneling event from the *LW* to the *UW* at the drain junction. The short source junction poses the dominant resistance in the circuit, and the measured conductance is determined by the tunnel coupling between the two wires at the source. We measure the differential conductance $\partial I/\partial V$ between the source and the drain using standard Lock-in technique. Typically *dV* is *14μV* and the modulation frequency is a few Hz. Measurements are taken in a dilution refrigerator at T=50-500mK. The reported results are insensitive to the temperature in this range.

Due to conservation of energy and momentum in the tunneling process [11], tunneling is only allowed when occupied states from the *UW* dispersion overlap, in momentum and energy, unoccupied states from the *LW*, or vice versa. We use this selective tunneling process to find the dispersion structure of the electrons in both wires. A bias *V* applied between the source and drain junctions generates a relative energy shift of *-eV* between the two wires dispersions. A magnetic field $B^X$ applied perpendicular to the wires plane boosts the tunneling electrons, and creates a momentum shift given by $-edB^X$ between the two dispersions, with *d* the distance between the wires centers. When the momentum and energy shift between the dispersions places the Fermi point of one wire on the dispersion of the other wire (Fig. 1c), a strong differential conductance signal is obtained, marking the onset of a tunneling channel between the wires. Thus, each Fermi point in one wire maps out the entire dispersion structure of the other wire (more details about this measurement technique can be found in [11]). Figure 1d presents the result of such a measurement. In this case, the *LW* has five occupied 1D modes, while the *UW* has two. The *LW* modes, scanned by a Fermi point of the *UW* dispersion, are clearly shown as five parabolic bands (see Fig. 1e). In addition, we identify the *UW* dispersions as scanned by the Fermi points of the LW modes, shown as inverted parabolas (these have a relatively high density, so the curvature near the Fermi level is indistinct). Although momentum and energy conservation allows every pair of modes to scan each other, several such pairs are missing in the measurement, as their tunnel-coupling is reduced by small overlaps between the wavefunctions (i.e. orthogonality) [11]. For example, the feature associated with tunneling between the second *LW* mode and the first *UW* mode is (almost) indiscernible in the measurement.

The electrons in the 1D channels are tightly confined in the growth (*z*) direction by atomically defined planes. Thus, the magnetic field $B^X$ has little influence on the electronic wavefunction. In the $\hat{x}$ direction, confinement is atomically sharp on the cleave side, but defined by electrostatic attraction to the cleave dopant layer on the opposite direction (Fig. 1a). This softer confinement is further screened by the electrons in the wire, and allows some freedom in the electronic distribution along the cross section of the wire when a magnetic field component $B^Z$ is applied. When $B^Z=0$, a system that has negligible thickness in the *z* direction will be invariant under *(y→-y, z→-z)*, and as a result the measured dispersion structure is symmetric to $B^X \rightarrow -B^X$, or correspondingly $k_y \rightarrow -k_y$. This symmetry is violated once a perpendicular magnetic field component $B^Z$ is applied, rendering the dispersion asymmetric. Electrons with momentum $\hbar k_y > 0$ are pushed by the Lorentz force towards the sharp edge, causing the dispersion to become steeper. Conversely, electrons with momentum $\hbar k_y < 0$ are pushed towards the soft confinement leading to a flattening of the dispersion, as well as a spatial separation between electrons of different momenta.

We focus on the soft side dispersion in the *LW*, as mapped by the Fermi point of the sharp side in the ground state of the *UW*. This Fermi point is well separated in momentum from other parts of the *US* dispersion, avoiding possible complications arising from a nontrivial charge structure in the *US*. The sharp side dispersion of the *LW*, as well as the electron structure of the upper system, are discussed elsewhere [12]. Separate control of the magnetic field component $B^Z$ affecting the electronic structure, and $B^X$ used to perform the spectroscopy, is realized by placing the sample on an *in situ* rotating stage.

Figure 2(a-f) presents the measured dispersion structure for several values of $B^Z$. A Hartree-Fock calculation of the expected signal is presented in Fig. 2(g-l), where for simplicity of the current discussion only transitions involving tunneling between the *UW* lowest mode and the *LW* are presented. For $B^Z=0.7T$ (Fig. 2a), no significant change is observed in the dispersions of the lowest modes. As stated, the dispersion line of the second *LW* mode is not visible in both measurement and calculation (Fig. 2g) due to orthogonality between the first mode wavefunctions in the *UW* and second mode wavefunctions in the *LW*. As $B^Z$ is increased, the soft-side dispersion in the *LW* is strongly modified while the sharp-side *UW* dispersion in weakly affected, invalidating this orthogonality. Consequently, for $B^Z=2T$ (Fig. 2b), the 2$^{nd}$ orbital mode of the *LW* is visible crossing the first mode dispersion ($B^X \approx 6.5T$ in Fig. 2b and the calculated Fig. 2h). This crossing stems from the different response of the two modes to the applied magnetic field. The first mode is strongly confined by the edge well, and retains its parabolic dispersion structure. The second (higher energy) mode is less tightly confined. As stated, the application of a magnetic field smoothes the $\hbar k_y<0$ part of the dispersion, leading to a crossing with the first mode. Another consequence of the application of $B^Z$ is an increase in the gap between the band bottoms of the 1$^{st}$ and 2$^{nd}$ modes, due to the added magnetic confinement. Indeed, as $B^Z$ is increased the 2$^{nd}$ mode occupation is decreased as its dispersion rises in energy (Fig. 2c). For $B^Z \geq 2.4T$ an avoided-crossing structure develops between the 1$^{st}$ and 2$^{nd}$ mode. The hybridization of the 1$^{st}$ and 2$^{nd}$ modes at the intersection is enhanced as the applied field is increased, leading to a larger separation at the crossing. The appearance of flat features for $4T<B^X<5.5T$ is associated with the emergence of Landau levels in the *US* [12].

Up to this point, the qualitative picture describing the wire dispersion was essentially captured within a single-particle account (i.e., interaction effects were not required to qualitatively explain any of the observed features). However, a deviation from such simple description is observed for $B^Z \geq 2.4T$. Here, the soft side of the second mode dispersion ($B^X \approx 6.5T$ in Fig. 2d) is split near the Fermi level. We attribute this splitting to a Dempsey-type reconstruction, in which one spin type ('down' in Fig. 2d) becomes less occupied than the other. This picture is reproduced in the calculation (Fig. 2j). Since, as described, this momentum-space occupation picture is related to a spatial distribution of the different modes, the reconstruction is similarly expressed as a development of a spin polarized strip in the cross section of the wire. As $B^Z$ is increased this splitting grows (Fig. 2d-f), representing a broadening of the spin polarized strip. This behavior was reproduced in several samples during many cooldowns, and was not affected by temperature in the measured range. For $B^Z>4.4T$ the entire 2$^{nd}$ mode becomes spin polarized (Fig. 2f and the calculated 2l), as one spin type is completely depleted. Quantitative discrepancies between the calculation and the measurements, most notably a difference in the energy scales, are related to assumptions used in the Hartree-Fock model [12]. Interestingly, the model predicts that the first wire mode is similarly spin polarized for $B^Z>1T$, as evident in the splitting between the dispersions of the two spin types in Fig. 2j-k. However, orthogonality of the spin split band and the probing *UW* wavefunction renders this structure invisible to the tunneling conductance measurement.

As stated, the phase transition into a spin polarized edge is controlled by the competition between the confining potential and the Hartree *e-e* interaction. Following Dempsey *et al.*, we use the dimensionless measure of this competition defined as

$$\alpha = \frac{l_B \partial V_{Conf.}/\partial x}{e^2/\varepsilon l_B}. \tag{1}$$

Here, $l_B$ is the magnetic length, $V_{Conf.}$ is the confinement potential and $\varepsilon$ the dielectric constant. The numerator measures the energy cost involved in creating a polarized strip of typical width $l_B$, while the denominator gives an energy scale to the corresponding gain in Coulomb energy. Although we have no handle on the confining potential slope, we can probe different values of $\alpha$ by controlling the applied magnetic field $B^Z$. Figure 3 presents the dependence of the polarized strip width on $\alpha$. A single parameter, namely the confining potential slope $\partial V_{Conf}/\partial x$, was deduced from the numerical model, by requiring a good agreement between the calculated and measured spin strip width as function of $B^Z$. We find the transition to a spin polarized edge at $\alpha \approx 5$, where for larger values (smaller magnetic fields, or equivalently sharper confinement) no spin polarization is obtained. The polarized strip width is found to be comparable to a magnetic length. The inset to Fig. 3 presents the predicted phase transition diagram for an edge state of a QHE, where a similar structure is expected, although the specific value of $\alpha$ where the transition is obtained ($\alpha \approx 0.4$) is different. This difference is predominantly related to the fact the our results discuss the development of a spin polarized strip in the second mode, while Dempsey *et al.* consider the phase transition of the lowest mode. Studying the predicted transition using the Hartree-Fock calculation, we find that when the conditions are such that the wire has only one populated mode, the spin polarized strip arises for $\alpha \approx 0.2$. The remaining discrepancy is reasonable, considering the different assumptions about the confinement potential and screening in both cases.

In conclusion, we have measured the dispersion structure of a quasi-1D wire under strong magnetic fields. We observed a splitting of the 2$^{nd}$ band and identified it as a spin reconstruction, in which a spin polarized strip is created along the cross section of the wire through a second order phase transition, similar to the spin reconstruction predicted for QHE edge states.

We would like to thank C. Chamon for helpful discussions. This work is supported by the NSF under contract number DMR-0707484 and DMR-0906475 and by Microsoft Corporation Project Q.

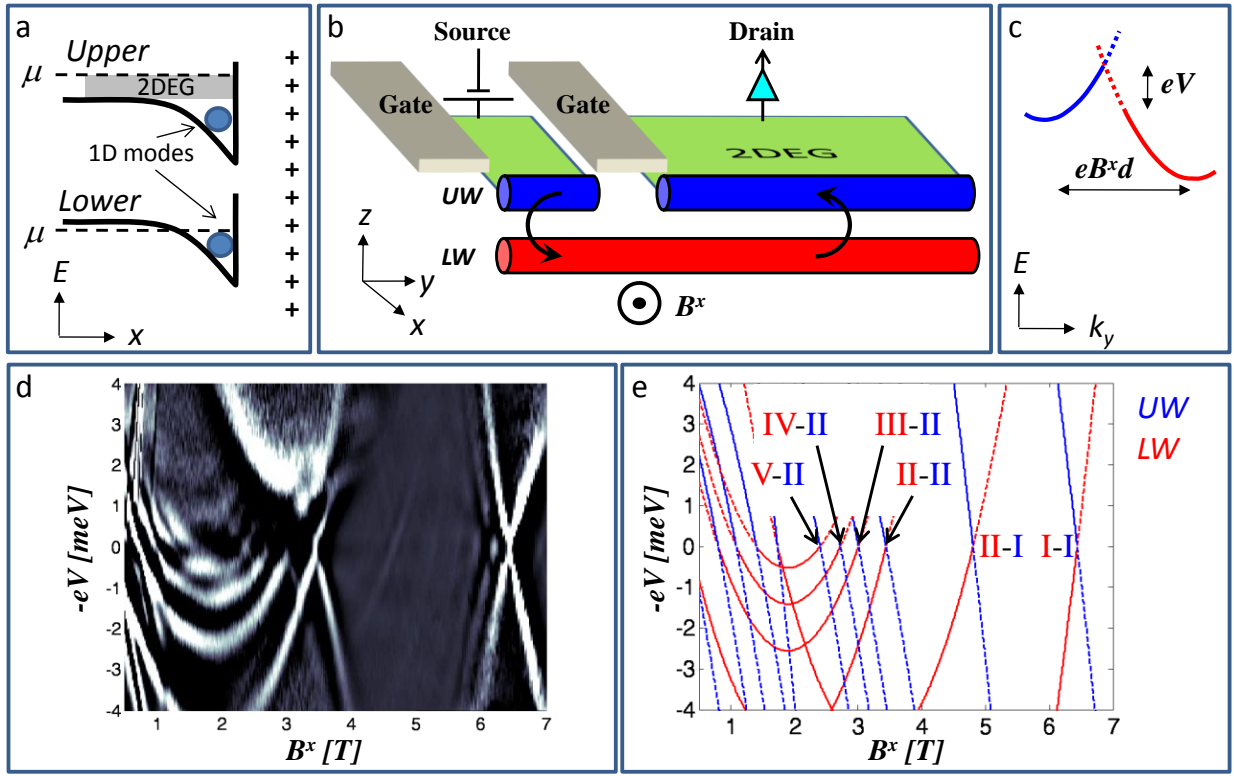

Figure 1: (a) The potential profile in the CEO double-wire system. The dopant layer added on the cleaved edge gives rise to an attractive triangular potential well, creating confined 1D states. In the upper system (*US*), the upper wire (*UW*) resides at the edge of a populated *2DEG*, used to make contact with the leads. The lower wire (*LW*) resides at the edge of an unpopulated *2DEG*. (b) The measurement setup. One top-gate (left) locally depletes both systems below it, while another top gate (right) depletes only the *US*, thus defining a short source junction and a long drain junction. Transport takes place by tunneling from the *US* to the *LW* at the source junction, and a second tunneling event from the *LW* to the *US* at the drains. (c) The technique of momentum resolved tunneling. The relative position of the *US* and *LW* dispersions in energy and momentum is controlled by an applied bias $V$ and magnetic field $B^x$. (d) A $B^z=0$ measurement. *LW* dispersions appear as concave parabolas, *UW* as convex parabolas. (e) Identification of the different modes constituting the observed dispersion structure.

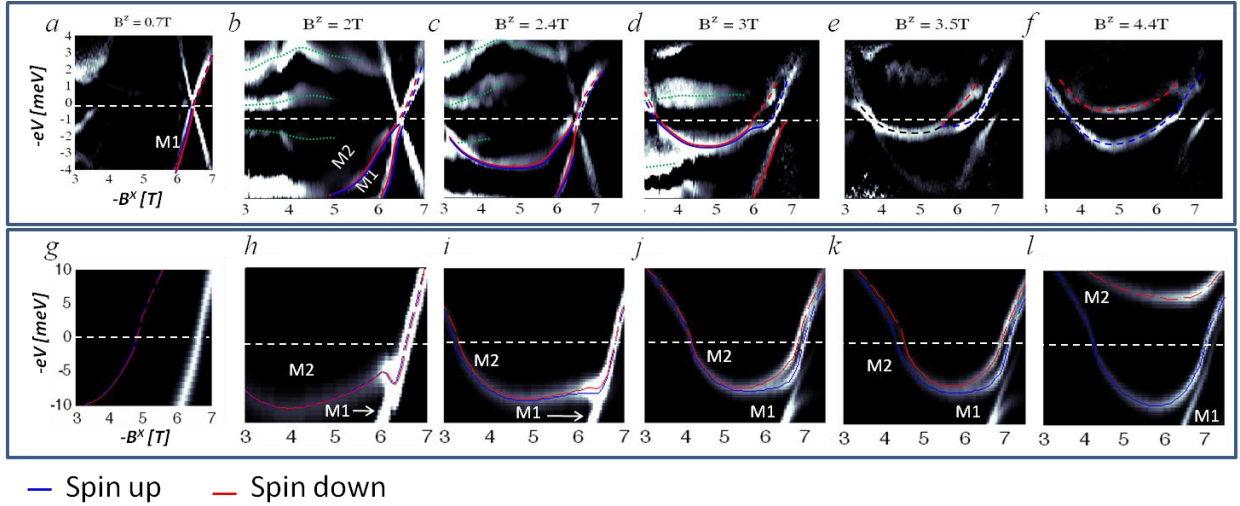

— Spin up    — Spin down

Figure 2: Wire dispersion structure under perpendicular magnetic field. (a-f) Measured dispersions. An avoided crossing between mode 1 and 2 in the wire is developed for $B^Z>2T$. The two spin orientations of the second mode are traced in blue and red. While for weak $B^Z$ the 2nd mode is spin degenerate, for $B^Z>2.4T$ a spin polarized strip is developed in the cross section of the wire. As $B^Z$ is increased this strip broadens, and for $B^Z>4.4T$ the wire becomes fully spin polarized. Horizontal features (green dotted lines) originate from direct tunneling to Landau levels in the *US*. (g-l) Hartree-Fock prediction for the measured dispersions, in the *LW* as scanned by momentum resolved tunneling from the 1$^{st}$ mode in the *UW*. Brightness of the calculated features is determined by the tunneling efficiency. The calculated dispersions reproduce the emergence of the spin polarized strip in the 2nd wire mode. In addition, the calculation predicts a similar polarization for the 1st mode, which is hard to discern (and invisible in the measurement) due to tunneling orthogonality.

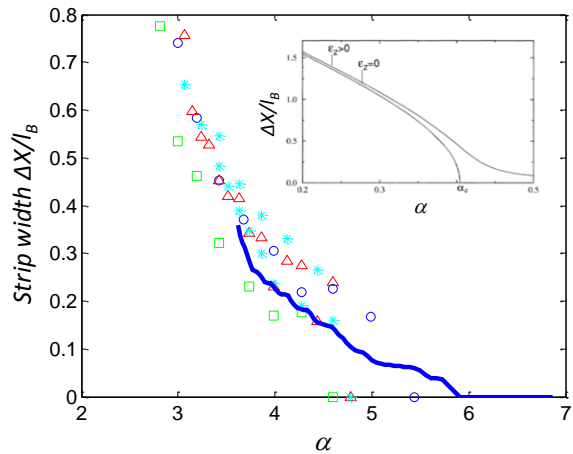

Figure 3: Dependence of the (normalized) spin polarized strip width on the dimensionless parameter $\alpha$. Different symbols correspond to different cooldowns. Results of the HF calculation presented in solid line. The polarized strip appears through a 2$^{nd}$ order phase transition at $\alpha \approx 5$, and becomes broader as $\alpha$ decreases. (Inset) Predicted polarized strip width for a QHE edge state (taken from Dempsey *et al.* [3]). The lower curve shows the behavior disregarding the Zeeman energy, while the upper curve includes a Zeeman splitting corresponding to $B=2T$ in GaAs.